\documentclass{phostproc}

\newcommand{\derivp} [2] {\frac {\partial #1 } {\partial #2} }

\title{A simple representation of oscillation modes in stars:\\from mixed modes coupling to glitches}
\author{Charly Pinçon$^{1,2,3}$}

\affiliation{$^{1}$  Institut d’Astrophysique Spatiale, Univ. Paris-Sud, CNRS, Universit\'e Paris-Saclay,
B\^atiment 121, 91405 Orsay CEDEX, France\\
			 $^{2}$  STAR Institute, Université de Liège, 19C Allée du 6 Août, B-4000 Liège, Belgium\\
			 $^{3}$  LESIA, Observatoire de Paris, PSL Research University, CNRS, Sorbonne Universit\'es, Univ. Paris Diderot, 5 place Jules Janssen, 92195 Meudon, France}

\shorttitle{Physical representation of bi-cavity oscillation modes}
\shortauthors{Charly Pinçon}

\abs{Analytical resonance conditions for oscillation modes in stars are very helpful both to predict and to examine their frequency spectra, as well as to make the link with their internal properties. In this short paper, we introduce a general quantization expression for oscillation modes accounting for the possible existence of a local sharp variation in the equilibrium structure, a so-called glitch. This representation is based on a direct adaptation of the progressive-wave picture of mixed modes proposed by \cite{Takata2016b}. In this formulation, a glitch turns out to be characterized by three parameters: its acoustic depth, the phase lags introduced after the wave reflection at the considered point, and a coupling factor. Such an expression has two main advantages. First, it can be easily applicable to a lot of different structural configurations. Second, it does not assume that the glitch is a small perturbation. Actually, we check that the obtained expression tends to the formulations previously derived when the glitch is weak. These research notes represent a preliminary step towards a more generalized description of multi-cavity oscillation modes, that was briefly addressed in the poster presented at the {\it PHOST} conference.}

\begin{document}

\maketitle

\section{Introduction}

In a general picture, gravito-acoustic waves in slowly-rotating stars can travel back and forth several times between the center and the surface where they are reflected. The trapping and the constructive interferences of such progressive waves may then result in global oscillation modes with a discrete frequency spectrum. The role of asteroseismology is thus to convert this set of eigenfrequencies into information on the properties of stars.

Different vibrational configurations are in theory possible in the interior of low-mass stars \citep[e.g.,][]{Unno1989}. For instance, in the Sun, very high-frequency oscillations are expected to form acoustic modes propagating through a single cavity located in the external envelope, or P cavity. In contrast, very low-frequency oscillations are expected to form gravity modes propagating in the inner radiative region, or G cavity. In an intermediate range of frequencies, another type of configuration can exist in which modes can oscillate in two distinct cavities -- the inner G cavity, where they behave as gravity modes, and the external P cavity, where they behave as pressure modes -- both separated by an intermediate barrier where modes have an evanescent behavior and are partially reflected/transmitted. These are the so-called mixed modes with a dual pressure-gravity character, responsible for the occurrence of avoided-crossings during stellar evolution \citep[e.g.,][]{Aizenman1977,Shibahashi1979}.

Although unobservable in main-sequence stars, mixed modes could be detected in the frequency spectrum of thousands of red giant stars observed by the satellites CoRoT \citep[e.g.,][]{Baglin2006a,Baglin2006b} and {\it Kepler} \citep[e.g.,][]{Borucki2010}. The analysis of the frequency pattern of mixed modes then provided a lot of constraints not only into the outer layers of these stars, but also into their innermost ones \citep[e.g.,][and references therein]{Hekker2017}. In particular, the physical intepretation of the data was partly made possible through the exploitation of the asymptotic expressions of mixed modes obtained by \cite{Shibahashi1979} and \cite{Tassoul1980} and, later, \cite{Takata2016a}.
Actually, \cite{Takata2016b} showed that these latter quantization conditions follow a unique and general analytical form relying on basic physical principles.
Such a simple expression has the advantage to highlight the parameters of interest associated with mixed modes and provides a practical tool to extract the physical information from real spectra.

This physical representation is not restricted to mixed modes and can also be adapted to any modes propagating in two distinct cavities, coupled by a given intermediate barrier. This gives rise to the idea of applying it to the case of glitches. A glitch denotes the perturbation of the frequency pattern induced by a sharp variation in the stellar structure, that is on a lengthscale smaller than the oscillation wavelength \citep[e.g.,][]{Vorontsov1988,Mazumdar2014}. Close to such a region, the WKB description of oscillations is not valid and partial wave reflection must occur. This is analogous to what happens close to the evanescent region of mixed modes. In this short study, we thus aimed at adapting the original formulation of mixed modes by \cite{Takata2016b} to the case of glitches. In Sect.~\ref{takata}, 
we briefly introduce the expression of mixed modes provided by \cite{Takata2016b}. Its adaptation to glitches is then developed and discussed in Sect.~\ref{glitch}. Preliminary conclusions are finally formulated \smash{in Sect.~\ref{concl}.}

\section{General formulation of mixed modes}
\label{takata}

\citet[][see Fig.~3 of this paper]{Takata2016b} represented mixed modes in a very general way. This formulation relies on two main assumptions. First, it considers the case of slowly rotating stars, so that the problem is spherical (i.e., in the radial direction). In this picture, the G and P resonant cavities are unidimensional and located between the radii $r_1$ and $r_2$, and $r_3$ and $r_4$, respectively, such as $r_1 < r_2 < r_3 < r_4$. Second, it assumes that the WKB is met in both the G and P resonant cavities, meaning that the oscillation wavelength is much smaller than the variation scale height of the medium \citep[e.g.,][]{Gough2007}. In other words, the wavefunction in a given $J$ cavity, which we denote $\varphi_J(r)$ with $r$ the radius in the star, can be written as the sum of a progressive and a regressive plane waves both multiplied by a constant (complex) amplitude, that is in the form of
\begin{equation}
\varphi _J (r) = \underbrace{a_{J,+} ~e^{+ ix_J(r)}}_{{\rm Upward~energy~ray}}+\underbrace{a_{J,-} ~e^{- ix_J(r)} }_{{\rm Downward~energy~ray}}\; .
\label{wf}
\end{equation}
Such a form is still possible with an appropriate change of variable in the WKB approximation \citep[e.g.][]{Shibahashi1979,Tassoul1980,Takata2016a}. In the latter expression, the $x_J$ coordinate denotes the phase function in a given $J$ cavity, that is in general defined as a linear function of the wavenumber integral with respect to radius. We emphasize that its definition can be totally different from a cavity to another one. In Takata's formulation, the direction of propagation that is considered is the one of the wave energy, i.e. of the group velocity. By convention, the subscript + (-) in Eq.~(\ref{wf}) is related to an energy ray propagating upwards (downwards). Note also that the time-dependence of the wavefunction is assumed to vary as \smash{$e^{-2 i \pi \nu t}$}, with $\nu$ the oscillation frequency. As a consequence, $x_J$ must increase with radius if the group velocity is in the same direction as the phase velocity. In other words, the upward (downward) energy ray in Eq.~(\ref{wf}) can be identified as the progressive (regressive) component. In contrast, $x_J$ must decrease with radius if the group velocity is in the opposite direction of the phase velocity. Equivalently, it means that the downward (upward) energy ray in Eq.~(\ref{wf}) can be identified as  the progressive (regressive) component.

Finally, the intermediate evanescent region is represented as a barrier located between $r_2$ and $r_3$, and characterized both by a reflection coefficient for the amplitude, denoted $R$, and by a phase lag introduced at the reflection\footnote{The phase lags at reflection here correspond to the amplitude ratio of the incident plane wave to the reflected one, in the sense of the group velocity.}, denoted $\delta$, of an upward incident energy ray coming from the inner G cavity\footnote{As shown by \cite{Takata2016b}, the reflection of a downward incident energy ray coming from the external (P) cavity on the intermediate barrier is associated with with a phase lag and a reflection coefficient equal to $\pi-\delta$ and $R$, respectively, so that no additional parameter needs to be introduced to fully describe the wave reflection-transmission problem.}. For boundary conditions close to the center and the surface, the wave reflection is supposed to be total and the possible associated phase lags introduced after reflection (i.e., at $r_1$ and $r_4$) are denoted $\theta_C$ and $\theta_S$, respectively.

In this framework, using basic wave principles (i.e., time-shift and time-reversal symmetries, superposition principle and energy conservation), \cite{Takata2016b} demonstrated that the resonance condition reads 
\begin{equation}
\cot \Psi_{\rm G} \tan \Psi_{\rm P} = \frac{1-R}{1+R} \equiv q \; ,
\label{resonance}
\end{equation}
where $q$ is the so-called mixed mode coupling factor and where the phase terms are provided by
\begin{eqnarray}
\Psi_{\rm G} &=\displaystyle x_{\rm G}(r_1) - \frac{\theta_{\rm C}}{2} - \frac{\delta }{2} +\frac{\pi }{2} \label{phase g}\\
\Psi_{\rm P}&= \displaystyle x_{\rm P}(r_4)+ \frac{\theta_{\rm S}}{2} - \frac{\delta }{2} +\frac{\pi }{2} \; .
\label{phase p}
\end{eqnarray}
Since the group velocity is in the opposite (same) direction as the phase velocity, the phase function must increase downward (upward) in the G cavity (P cavity)\footnote{Indeed, the radial wavenumber is (inversely) proportional to the oscillation frequency for asymptotic (gravity) pressure modes.}. For sake of simplicity, we assume in all the following that there is no phase lag introduced after the wave transmission through the intermediate barrier (i.e., the evanescent region in the case of mixed modes). Therefore, the values of the $x_{\rm G}$ and $x_{\rm P}$ coordinates at the lower and the upper boundaries of the evanescent region, respectively, must be equal, or equivalently, \smash{$x_{\rm G}(r_2)=x_{\rm P}(r_3)$}. It is thus possible to fix the origin of both coordinates at these points, such as we can write
\begin{eqnarray}
x_{\rm G}(r) &= \displaystyle\int_r^{r_2} k_r {\rm d}r \\
x_{\rm P}(r) &= \displaystyle\int_{r_3}^{r} k_r {\rm d}r  \; ,
\end{eqnarray}
where $k_r$ is the local radial wavenumber. In these considerations, the values of $ x_{\rm G}(r_1)$ and $x_{\rm P}(r_4)$ in Eqs.~(\ref{phase g}) and (\ref{phase p}) are positive and must be identified as
\begin{eqnarray}
x_{\rm G}(r_1)&=\displaystyle \int_{r_1}^{r_2} k_r {\rm d}r \\
x_{\rm P}(r_4)&=\displaystyle \int_{r_3}^{r_4} k_r {\rm d}r  \; .
\end{eqnarray}
This representation generalizes the asymptotic expressions obtained by \cite{Shibahashi1979} and \cite{Tassoul1980} in the limiting case of a very thick evanescent zone and by \cite{Takata2016a} in the other limiting case of a very thin evanescent region.

\section{Extension to glitches}
\label{glitch}

Glitches define the modification of the mode frequencies induced by the presence of a sharp feature in the equilibrium structure compared to the case where the variation of the structure remains smooth, that is on a lengthscale much larger than the oscillation wavelength. In the following, we assume that such a rapid variation locally exists at a given radius $r_g$. This steep gradient in the structure can be assimilated to a barrier where incident waves are partially reflected and transmitted. In the framework of the representation of \cite{Takata2016b}, we characterize it by a reflection coefficient, $R_g$, and a phase lag at reflection, $\delta_g$. In this section, we aim at analytically describing such a configuration from a very general point of view.

\subsection{Adaptation of Takata's general picture}

Such a configuration is actually very similar to mixed modes. It is composed of two resonant cavities, the inner one located between $r_1$ and $r_g$ and denoted the I cavity, and the external one located between $r_g$ and $r_{\rm 4}$ and denoted the E cavity (i.e., such as $r_1 < r_g <r_4$). Only two minor differences exist between both cases. First, the region associated with the glitch is very thin and assumed to be localized in one given layer. In other words, it would be equivalent to assume $r_3=r_4=r_g$ in the case of mixed modes. Second, since the intermediate barrier is not an evanescent region in the case of glitches, the modes behave in a similar way in each cavity, that is either as pressure modes or as gravity modes. This means that the phase function associated with each cavity, denoted $x_{\rm I}(r)$ and $x_{\rm E}(r)$, must increase in the same direction.

By default, we consider that the phase and the group velocities are in the same direction (i.e., as pressure modes), so that the phase functions increase with radius. Using similar assumptions to  those in Sect.~\ref{takata}, we consider that the origin of the phase functions in both cavities is at $r_g$, so that \smash{$x_{\rm I}(r_g)=x_{\rm E} (r_g) = 0$}. As a result, we obtain
\begin{eqnarray}
x_{\rm I}(r) &= -\displaystyle\int_r^{r_g} k_r {\rm d}r \label{x_I} \\
x_{\rm E}(r) &= \displaystyle\int_{r_g}^{r} k_r {\rm d}r  \label{x_E}\; .
\end{eqnarray}
Given the similarity with the representation of mixed modes provided by \cite{Takata2016b}, the resonance condition accounting for a glitch can therefore be deduced from \smash{Eqs.~(\ref{resonance})-(\ref{phase p})} and Eqs.~(\ref{x_I})-(\ref{x_E}) by making the coordinate substitution $x_{\rm G} \leftarrow x_{\rm I}$ and $x_{\rm P} \leftarrow x_{\rm E}$, so that
\begin{equation}
\cot \Psi_{\rm I} \tan \Psi_{\rm E} = - \frac{1-R_g}{1+R_g} \equiv - q_g \; ,
\label{resonance g}
\end{equation}
where $q_g$ is the coupling factor associated with the glitch and the phase terms are equal to
\begin{eqnarray}
\Psi_{\rm I} &=\displaystyle \int_{r_1}^{r_g} k_r {\rm d}r + \frac{\theta_{\rm C}}{2} + \frac{\delta_g }{2} +\frac{\pi }{2} \label{phase I}\\
\Psi_{\rm E}&= \displaystyle \int_{r_g}^{r_4} k_r {\rm d}r + \frac{\theta_{\rm S}}{2} - \frac{\delta_g }{2} +\frac{\pi }{2} \label{phase E}\; .
\end{eqnarray}
Using trigonometric formulas, Eq.~(\ref{resonance g}) can also be rewritten
\begin{eqnarray}
\sin( \Psi_{\rm I}+\Psi_{\rm E}) = R_g \sin(\Psi_{\rm I} - \Psi_{\rm E} )\; .
\label{resonance 2}
\end{eqnarray}
In this physical representation, a glitch is thus characterized by a coupling factor, its position in the cavity and the phase lags introduced at reflection. The analytical resonance condition given in Eqs.~(\ref{resonance g})-(\ref{phase E}) is general and do not assume that the impact of a glitch on the mode frequencies is weak, as usually done in previous formulations (see Sect.~\ref{weak}). 

In the case where the phase and the group velocities are in opposite directions (i.e., as gravity modes), similar relations can be obtained. In this case, the phase functions in the I and E cavities decrease with respect to $r$. As a consequence, the resonance condition is also provided by \smash{Eqs.~(\ref{resonance g})-(\ref{phase E})}, except that we must apply the substitution $\theta_{\rm G} \leftarrow -\theta_{\rm G}$, $\theta_{\rm S} \leftarrow -\theta_{\rm S}$ and $\delta _g \leftarrow - \delta_g$. At this point, we note that we retrieve a similar form as the one found by \cite{Brassard1992} who studied glitches in the frequency pattern of gravity modes in ZZ Ceti stars. In these stars, glitches may result from the sharp gradient in the chemical composition near the hydrogen-burning shell. \cite{Brassard1992} modeled the induced rapid variation in the Brunt-Väisälä frequency by a step function, which was assumed to discontinuously change from a value $N_-$ to a value $N_+$ (i.e., such as $N_+ < N_-$). In this special case, the comparison between the resonance condition that they obtained and Eqs.~(\ref{resonance g})-(\ref{phase E}) enables us to make the identification $q_g=(N_+/N_-)$ and $\delta_g=0$. To be complete, we also notice that a similar relation was derived later by \cite{Miglio2008}, who studied glitches in SPB and $\gamma$ Doradus stars.

\subsection{Usual case without a glitch}

\label{usual}

When the variation of the equilibrium structure is smooth and hence $R_g =0$, the resonance condition in Eq.~(\ref{resonance 2}) for pressure modes (or, in a more general way, when the phase and the group velocities are in the same direction) results in
\begin{equation}
\Psi_{\rm I} + \Psi_{\rm E} = \displaystyle\int_{r_1}^{r_4} k_r {\rm d}r +\frac{\theta_{\rm C} + \theta_{\rm S}}{2} = n \pi \; ,
\label{no glitch}
\end{equation}
with $n$ a given integer. This condition is equivalent to the Bohr-Sommerfeld's quantization rule in quantum mechanics and represents a generalization of the asymptotic expressions for pressure modes that were derived by \cite{Shibahashi1979} and \cite{Tassoul1980}.

In the case of gravity modes, the same relation can be easily obtained, except that we still must make the substitution $\theta_{\rm G} \leftarrow -\theta_{\rm G}$ and $\theta_{\rm S} \leftarrow -\theta_{\rm S}$.

\subsection{Weak perturbation for pressure modes ($R_G\ll1$)}
\label{weak}

Previous formulations of glitches in the case of pressure modes usually considered that the modification of the eigenfrequencies compared to the smooth case (i.e., without a glitch) is small. In other words, they considered that the frequency perturbation is smaller than the frequency difference between two consecutive eigenmodes. In order to discuss this specific case in the framework of the present physical representation, we assume that $R_g \ll 1$ so that the eigenfrequencies are expected to be only slightly modified by the glitch. In these considerations, we can formally rewrite each eigenfrequency $\nu$ as
\begin{equation}
\nu=\nu_{0}+\delta\nu\; ,
\end{equation}
where $\nu_{0}$ is the value of the eigenfrequency in the smooth case and $\delta \nu$ is the glitch-induced frequency perturbation. To go further, we define the frequency-dependent function
\begin{equation}
 \Psi=\Psi_{\rm I}+\Psi_{\rm E} \; .
 \label{Psi}
 \end{equation}
According to Eq.~(\ref{no glitch}), we must have $\Psi(\nu_{0})=n \pi$ with $n$ the corresponding integer. In the weak perturbation limit, the perturbation of the phase $\Psi$ must be small, that is $(\partial \Psi/\partial\nu)_{\nu_0}  \ll \pi /\delta \nu $. At first-order, we can therefore write
\begin{equation}
\Psi(\nu) \approx n\pi+  \delta \nu \left( \derivp{\Psi}{\nu}\right)_{\nu_{0}} \; .
\label{Psi 1}
\end{equation}
A zeroth-order expansion of $\Psi_{\rm I}-\Psi_{\rm E}$ around $\nu_{0}$ also gives
\begin{equation}
 \Psi_{\rm I}(\nu)-\Psi_{\rm E}(\nu)= \displaystyle n\pi-2\Psi_{\rm E}(\nu_{0})  + \mathcal{O}\left[\delta \nu \left( \derivp{\Psi}{\nu}\right)_{\nu_{0}}\right]\; ,
 \label{Psi 0}
\end{equation}
where $(\mathcal{O})$ corresponds to the Bachmann-Landau's big O notation.
Injecting both Eqs.~(\ref{Psi})-(\ref{Psi 0}) in Eq.~(\ref{resonance 2}), the resonance condition at first-order for $R_g \ll 1$ and \smash{$(\partial \Psi/\partial\nu)_{\nu_0}  \ll \pi /\delta \nu $} leads to 
\begin{equation}
 \delta \nu \left( \derivp{\Psi}{\nu}\right)_{\nu_{0}}\approx - R_g \sin\left[2 \Psi_{\rm E}(\nu_{0})\right] \; .
 \label{first order}
\end{equation}
For pressure modes in the asymptotic limit, the radial wavenumber is provided in a good approximation by \citep[e.g.,][]{Shibahashi1979}
\begin{equation}
k_r \approx \frac{2 \pi \nu }{c_{\rm S}} \gg \frac{1}{r} \; ,
\end{equation}
where $c_{\rm S}$ is the sound speed. The $\Psi$ function is thus given at leading-order by the wavenumber integral over the I and E cavities. Indeed, the wavenumber integral represents the number of oscillation nodes in both cavities, which is very large in the asymptotic limit, while the phase lags $\theta_{\rm S}$ and $\theta_{\rm C}$ are of the order of unity only. We thus obtain
\begin{eqnarray}
\Psi(\nu) \approx \frac{\pi \nu}{\Delta \nu_0} \; ,
\end{eqnarray}
where $\Delta \nu_0$ is the large separation at frequency $\nu_0$ (i.e., we neglect the contribution of the frequency-dependence of $r_1$ and $r_4$ to the value of $\Psi$), which is defined as
\begin{equation}
\Delta \nu_0 = \left(2 \int_{r_{\rm 1}(\nu_0)}^{r_{\rm 4}(\nu_0)}\frac{{\rm d}r}{c_{\rm S}} \right)^{-1} \; .
\end{equation}
Similarly, the phase $\Psi_{\rm E}(\nu_0)$ in Eq.~(\ref{phase E}) is provided in a good approximation by
\begin{eqnarray}
\Psi_{\rm E}(\nu_0)\approx2\pi \nu_0\tau_{g}+\frac{\theta_{\rm S}-\delta_g}{2} + \frac{\pi}{2} \; ,
\end{eqnarray}
where $\tau_{g}$ is the acoustic depth of the region with the sharp gradient that is equal to
\begin{equation}
\tau_{g}=\int_{r_g}^{r_{\rm 4}(\nu_0)} \frac{{\rm d}r}{c}  \; .
\label{tau_g}
\end{equation}
At the end of the day, the glitch-induced modification of the mode frequency can be obtained from Eqs.~(\ref{first order})-(\ref{tau_g}) and reads
\begin{eqnarray}
\frac{\delta \nu}{\Delta \nu_{0} }\approx  \frac{R_g}{\pi} \sin\left( 4\pi \nu_{0}\tau_g+\theta_{\rm S}-\delta_g\right) \; .
\label{glitch final}
\end{eqnarray}
Equation~(\ref{glitch final}) has a similar form as the one used in previous works \citep[e.g.,][]{Monteiro2005,Mazumdar2014,Vrard2015}. It shows that a sharp variation or a discontinuity in the stellar structure leads to a sinusoidal signal in the frequency difference between two consecutive eigenmodes. The amplitude of the signal is directly proportional to the wave reflection coefficient at the considered region, which must depend in general on both the mode frequency $\nu_0$ and the amplitude of the sharp gradient. Its period as a function of the mode frequency is proportional to the acoustic radius of the glitch. Actually, this is true if and only if the wave phase lags at the surface and at the region associated with the glitch does not vary too much with the mode frequency $\nu_0$. If this condition is met, the phase offset is thus quasi constant and equal to $\theta_{\rm S} -\delta_g$. 
In the case of gravity modes, a similar expression can be found for the glitch-induced variation of the mode period in the weak perturbation hypothesis, as already shown for instance by \citet{Miglio2008} via the variational principle.

 \section{Concluding remarks}
\label{concl}

In this work, we adapted the physical formulation of mixed modes by \citet{Takata2016b} to the case of glitches. This new representation is general. It only assumes that the star is spherical (i.e., unidimensional) and that the WKB approximation is met in the resonant cavities where modes can propagate. The obtained expression for the glitch depends on three main physical parameters: the depth of the barrier associated with the glitch (i.e., with a sharp gradient in the equilibrium structure), a coupling factor related to the wave reflection coefficient at the barrier and measuring the degree of interaction between both cavities located on both sides on this latter, as well as the phase lags introduced at reflection/transmission. We demonstrated that the usual mode quantization relations in the case without a glitch and in the weak perturbation hypothesis can be retrieved from this general expression.

Such a simple analytical relation provides a useful tool to disentangle and characterize the physical information brought by glitches from real oscillation spectra. To go further, the link between the associated parameters and the internal structure can be subsequently made by more detailed analyses in the neighborhood of the barrier using asymptotic methods or simplified modeling, as done for instance in \cite{Brassard1992} or \cite{Miglio2008}. Such studies will also provide information on the possible frequency-dependence of the parameters over the observed frequency range and its impact on the interpretation of the measured values. To conclude, these notes are thus a first step towards a general formulation of more complex configurations accounting for a multitude of resonant cavities and barriers, as for instance in the case of buoyancy glitches in red giant stars \citep[e.g.,][]{Cunha2015}. This will be subject to a forthcoming paper (Pinçon {\it  et al.}, 2019, in prep.)

\section*{Acknowledgments} I am indebted to A. Noels, M.-A. Dupret and M. Farnir for their careful reading of the paper and their relevant comments. This work was partially supported by postdoctoral grants from Centre National de Recherche Scientifique (France) and F.R.S.-FNRS (Belgium).

\bibliographystyle{phostproc}
\bibliography{bib.bib}

\end{document}